\documentclass{svproc}
\usepackage{rotating}
\usepackage{url}

\usepackage{graphicx}
\usepackage{cite}
\usepackage{makecell}
\usepackage{multicol}
\usepackage{footmisc}
\usepackage{orcidlink}

\begin{document}
\mainmatter              
\title{Experimentation Practices in 
\\Indie Game Startups}
\titlerunning{Experimentation Practices}  

\author{Elizabeth Bjarnason\inst{1}\textsuperscript{, }\textsuperscript{\orcidlink{0000-0001-9070-0008}} \and 
Johan Linåker\inst{1} \textsuperscript{, }\textsuperscript{\orcidlink{0000-0001-9851-1404}} \and
Fabian Fagerholm\inst{2}\textsuperscript{, }\textsuperscript{\orcidlink{0000-0002-7298-3021}}}
\authorrunning{Bjarnason et al.}
%

\institute{Lund University, Dept. of Computer Science, Lund, Sweden\\ \email{\{elizabeth.bjarnason | johan.linaker\}@cs.lth.se} \and 
Aalto University, Helsinki, Finland \\
\email{\{fabian.fagerholm\}@aalto.fi}}

\maketitle              
\begin{abstract}
Small independent (indie) game startups face fierce competition when developing games with limited time, budget, and resources. Continuously testing and experimenting with new ideas, features, and content is essential to validate and guide development toward a viable and successful game. Such continuous experimentation (CE) requires access to users and user data, which is often very limited in the early stages of development, both in general and, in particular for indie game companies.
We provide insights and recommendations on how to perform experimentation and game testing in pre-release stages of indie game development
based on an exploratory interview survey with 10 independent game developers from different companies.  
We provide a framework for CE that highlights key parts to consider when planning and implementing an experiment. Current practices for experimentation within indie game development are also described for the early development stages of ideation, preproduction, feature production, and content production. 
Our empirical findings offer insights and recommendations into how to conduct experiments during pre-release stages of game development, specifically for Indie game companies. Other software startups, particularly those with a value proposition centred around interactive and satisfying user experiences, can also benefit from our research.

\end{abstract}

\section{Introduction}
The game industry is thriving with increasing yearly turnarounds and number of employees. For example, in Sweden the game industry represents 4 percent of the national gross service export with a yearly turnaround  of 86 billion SEK (including international subsidiaries) and a workforce of around 25,000 people globally, including 8,500 employees in 939 companies located in Sweden, of which a significant portion are small indie companies ~\cite{swedishgamesindustry2023developerindex}. The industry is fiercely competitive, and game development companies are pressured by short deadlines and need to stay cost-efficient\cite{schmalz2014risk}. Therefore, betting on a wrong idea or a feature that is not technically or commercially viable poses a significant business risk.

In the broader industry, software-intensive organisations have increasingly adopted Continuous Experimentation (CE) practices~\cite{ros2018continuous} to mitigate the risk of releasing products with weak market viability. With this experiment-driven development approach features and qualities are continuously evaluated based on user feedback~\cite{fagerholm2017right} with hypotheses defined upfront based on expected technical and/or commercial viability~\cite{auer2021controlled}. Ideally, experimentation should be performed throughout the entire game development process, from ideation and pre-release to post-launch. This approach guides development and product planning to create an engaging game experience that maximises the value creation~\cite{yaman2018continuous}. The pre-release stages are extra critical, as they include the majority of development activities. As development progresses, the cost and impact of pivoting increases.

CE requires access to user data to test hypotheses~\cite{fagerholm2017right}, which is especially challenging prior to release. At this stage, the number of users, and thus user data and feedback, is naturally limited~\cite{yaman2018continuous}. This challenge is especially taxing for gaming startups with limited resources and without the backing of larger game publishers. These companies, known as independent (\textit{indie}) game companies, are typically small and new, with limited resources, experience, and practice comparable to established game companies and studios~\cite{grabarczyk2016every}.  

In this chapter, we provide a framework of CE that highlights five key parts to consider when planning and implementing experimentation in game development based on our previous work.
We also outline how experimentation practices can vary over development stages from ideation to beta release. Our results are based on an exploratory semi-structured interview survey with 10 game developers from different indie game companies. We have synthesised our findings into a revised version of our framework of continuous experimentation and provide recommendations for experimentation practices within game development based on current practices as described by our interviewees. In the future, we aim to further explore experimentation practices through case studies of a wider range of companies including larger game companies and companies in other business domains, and to investigate how our recommendation can support companies in improving their development practices.

\section{State-of-the-Art}


While many CE-like activities can be observed in game development there is limited research of experimentation in the game industry and for indie game studies ~\cite{engstrom2020game} with some notable exceptions~\cite{yaman2018continuous, edison2023experimentation}.
Game analytics is a widely adopted practice within game development used for purposes such as in-game balancing, identifying bottlenecks in game levels, detecting bugs, reducing costs and risks, and negotiating with investors and publishers~\cite{su2021gameanalytics}.
Leveraging field data to inform decision-making is a fundamental aspect of CE.
Some specific aspects of CE have received direct attention in game development~\cite{engstrom2020game}, e.g., the exploratory use of prototyping in the requirements engineering process~\cite{edison2023experimentation}, and monetary feature-optimisation in post-deployment.
The area of User research also provides some input into how user feedback can guide game development~\cite{pagulayan2018applied}.
User research shares with CE the overall idea of basing development on an empirical approach.

Research describes a need to match experimentation practices to the different stages of game development~\cite{aleem2016game}.
Qualitative methods are primarily used in early development stages when the amount of quantitative user data is limited~\cite{yaman2018continuous}.
Play-testing is highlighted as the main overarching method, while A/B testing is common in the later post-production stages~\cite{hyrynsalmi2018minimum}.
Identifying and isolating features for evaluation is a challenge~\cite{koskenvoima2015small}, especially in the early development stages~\cite{jarvi2015lean}.
The evaluated feature set needs to be comprehensive enough to retain all essential aspects required for game-play, while still allow the experiment to assess the defined Key Performance Indicators (KPIs)~\cite{hyrynsalmi2018minimum}.

A distinction can also be noted in relation to the game genre and context.
For example, in the mobile game context and freemium-type games~\cite{koskenvoima2015small}, there is a strong belief in quantitative methods.
Andersen et al., demonstrate the potential of quantitative experimentation in guiding feature development in gaming~\cite{andersen2011on}. However, while this approach is applied within game development, in particular for mobile games, our research indicates that there are game companies where feature development tends to be opinion-based instead of based on the outcome of game testing.

Depending on the purpose and aspects being experimented on, different metrics are required, such as game mechanics (the rules of the game), game dynamics (what happens when the game is played), and aesthetics (the feeling of fun experienced by the player)~\cite{jarvi2015lean}.
When designing an experiment, each aspect requires specific consideration regarding the technical implementation and commercial viability~\cite{rosenfield2017mvps}.
Validation of aesthetics, or fun factor (regardless of technical or commercial viability) is specifically highlighted as a key challenge, due to the need to connect with a relevant target audience~\cite{schmalz2014risk}, while limiting the risk of leaking differentiating ideas and details to the public and potential competitors~\cite{aleem2016game}.
Therefore, play-testing is usually performed with internal developers or trusted associates, despite the risk of cognitive and confirmation biases~\cite{yaman2018continuous}.
Chueca et al. points to the need for new methods, tools, and processes to identify and validate requirements relating to the user experience of games~\cite{chueca2023consolidation}.


\section{Research Method}

Our findings are based on an interview survey 
of how CE is applied within indie game development prior to releasing a game. We  interviewed representatives from ten indie companies to gain a broad understanding of practices across different game startups. Seven of the startups have fewer than 5 employees, two have 8-9 employees, and one has over 50 employees.  Our interviewees' experience of game development ranges from four to fourteen years, and covers roles such as founder, co-founder,  game producer, game director, lead game developer, and lead game designer. Six of the interviewees were currently based in Sweden, two in Netherlands, one in Switzerland, and one in Ukraine.  
The interviews were performed online for about 30 minutes each and consisted of 
open questions regarding the use and experience of CE and related practices such as game user research, testing, and requirements engineering. Further details of our research method can be found in [anonymised].

\section{Continuous Experimentation Framework}
\label{sect:CEFramework}

\begin{figure}
    \centering
    \includegraphics[width=1\linewidth]{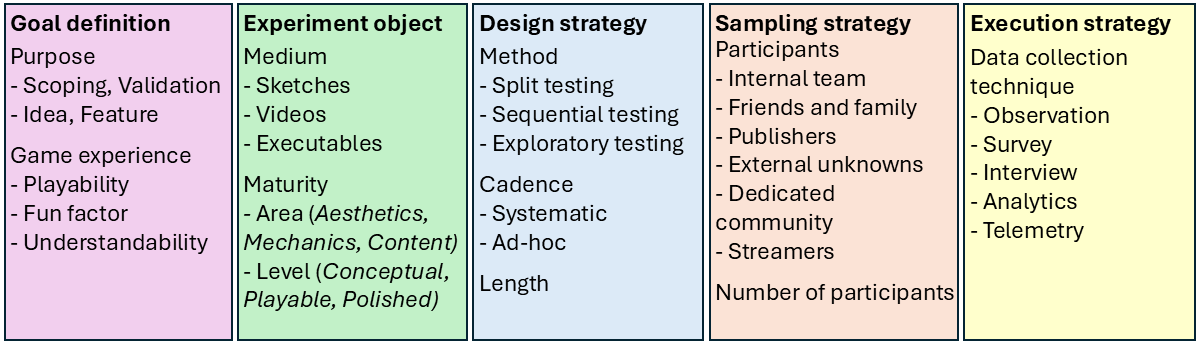}
    \caption{Overview of our framework for CE in game development including key parts of experimentation in the early stages of game development.}
    \label{fig:model}
\end{figure}

Our framework for CE consists of five key components that outline the rationale and methodology behind experimentation. These components are \textit{goal definition}, 
\textit{experiment object}, 
\textit{sampling strategy},
\textit{design strategy},  and \textit{execution strategy}, see Fig.~\ref{fig:model}. The current version of the framework includes some dimensions specific to game development, such as game experience. However, most of the dimensions are general, such as understandability, sketches and executables, and may apply to software development in any business domain.



\subsection{Goal Definition} 
The goal of experimentation consists of a combination of the \textbf{purpose} of an experiment and the aspect of the \textbf{game experience} that is being evaluated. The purpose concerns \textit{scoping} and \textit{validating} of \textit{game ideas} and \textit{features}. The \textit{game experience}~\cite{stahlke2018usertesting}, or user experience, that is evaluated concerns the  \textit{playability}, \textit{fun factor}, and \textit{understandability} experienced by the user while playing. Evaluation of these aspects is closely coupled to the design and development of the mechanics, aesthetics, and content of a game. For example, experimenting with \textit{``mechanics like weapons and items... [enables exploring] what kinds of choices players are making and if they feel that those are satisfying''}.
The importance of understandability varies depending on the type of game. For example, for a strategy game, the player needs to \textit{``learn... and understand the basic game mechanisms''} 
to play and enjoy the game and thus has a steeper learning curve compared to a first-shooter game.

\subsection{Experiment Object}

The experiment object refers to the artefact that is used to evaluate the game experience. The \textbf{medium} that is used (sketches, videos, or executable prototypes) and the aspect and level of \textbf{maturity} of the experiment object need to be aligned to the goals of the experiment. \textit{Sketches} of the art and imagined gameplay can be used to evaluate the overall game idea. \textit{Videos} demonstrating gameplay are often used to communicate new game ideas and concepts. \textit{Executable prototypes} are used to evaluate the actual game experience by observing participants playing the game. For this reason, experimenting with an executable prototype is often preferred even if this requires more time and effort than, e.g. sketching or prototyping a new idea in isolation. One interviewee motivated this preference with a desire to \textit{``get it [the change] into the game as quickly as possible ... to see if it works with the rest''}.
The maturity of the experiment object can vary regarding its \textit{level} of detail for different \textit{aspects} of the game, such as \textit{aesthetics}, \textit{mechanics}, and \textit{content}. The aesthetics relate to the graphical style and artwork of the game, while game mechanics concern the dynamics, features, and physical mechanics as experienced when playing a game. The content relates to game level design, and extending the game play with more levels and variation in using the mechanics. The maturity \textit{level} of the experiment object varies from \textit{conceptual} to \textit{playable} to a \textit{polished} representation of one or more of the aspects. Sketches that illustrate the aesthetics and mechanics of an envisioned game at a \textit{conceptual level} can be used to evaluate high-level game ideas and visions. In contrast, a rough executable prototypes with \textit{playable} \textit{mechanics} and \textit{aesthetics} can be used to demonstrate and evaluate the core mechanics and loop of a game. Similarly, as level design progresses, the \textit{content} is refined and expanded, and the overall game experience is evaluated through experimentation.

\subsection{Design Strategy} 

When designing and setting up experiments three main types of \textbf{methods} are used, namely \textit{split}, \textit{sequential}, and \textit{exploratory testing}, with variations in \textbf{cadence} and \textbf{length} of experiments. The choice of strategy depends on the experimentation goal, type of game, current development stage, and available resources. \textit{Split testing} involves comparing multiple versions and is often limited to two versions (i.e. A/B testing). This limitation for indie game development is due to the high cost of developing multiple variants of the game. \textit{Sequential testing} is the most common experimentation method used when developing features and content. In this method, the option considered to have the best potential is evaluated first, and if the feedback is negative, the alternative solution is implemented and evaluated. For example, when improving on functionality that is not working, developers  \textit{``just test the new version and see if the reactions have improved.''}
\textit{Exploratory testing} is a less formal experimentation method where there is no predefined hypothesis that is used either for a general and open evaluation, or to focus on a specific aspect such as game mechanics or fun factor. One game developer describe this as \textit{``If we have something that we know we want to know more about, then it is really focused... if it is more ...how does the game play experience feel ... then it is more open.''} 
The \textit{cadence} with which experiments are performed varies through \textit{structured systematic experimentation} to \textit{ad-hoc experimentation} with a \textit{length} commonly ranging between a weekend to a week. Sometimes longer experiments, spanning up to several weeks, are performed though updates during this period makes it hard to draw clear conclusions from the experiment since the feedback \textit{``at the beginning of the playtest and ... at the end of the playtest is actually feedback on two different builds''}.

\subsection{Sampling Strategy} 










Participants of an experiment are selected, or sampled, to represent a certain population, which involves selecting the \textbf{demographics} and \textbf{number of participants}. Indie game companies engage various types of participants, including their \textit{internal team} and \textit{publishers}, \textit{friends and family}, \textit{external unknowns}, and \textit{dedicated communities}. \textit{Streamers}, i.e. individuals who play and comment on games while simultaneously recording and sending live, are also included. Other game developers can provide \textit{``feedback [that is] even more useful ... [since] they have some development knowledge''} and 
tend to \textit{``see things that regular gamers do not think about''}.
The type and number of participants used typically depend on the current stage of game development and the availability of participants. The sampling also needs to align with the type of experiment (goals, design strategy, and experiment object). For physical playtests, indie game developers generally are satisfied with 8-10 individuals. For online and external playtests with players from the community or unknown externals, a larger number of participants are often used. For example, A/B test can be performed within a dedicated community of around 60-70 participants. 
Valuable insights can be gained from as little as \textit{``10 players [which can] often get you... [a] visible micro trend''}.
In other cases, 100 testers or more testers may be required for the input to be significant and trustworthy. 
However, indie game companies face major challenges in recruiting larger numbers of participants, 
and in attaining un-biased participants with a good representativeness of the target audience.  
For example, friends and family often have a positive bias and are pre-disposed to approve of the company's work. So, the representativeness of those selected for the experiment needs to be considered to mitigate the risk of biases, inconclusive, and misleading results.

\subsection{Execution Strategy}




Feedback is gathered from experiments through \textit{data collection techniques} such as \textit{observations}, \textit{surveys}, \textit{interviews}, and \textit{telemetry data}. Indie game companies commonly use \textbf{observations} of participants performing game play to gain insights into how the game is played, and if participants get stuck or do not understand certain aspects of the game. Often the screen is recorded to allow for later analysis and also enables recruiting participants irrespective of physical location. Watching people play can also be useful in mitigating biases, for example, in friends \& family, or when game players know that they are talking to the developer and are less likely to criticise. When observing participants during live game play testing, it is important not to interfere by asking leading questions or helping them, since this may influence their behaviour and responses.
Follow-up \textit{surveys} and \textit{interviews} are commonly used to complement observations. Surveys can consist of a mix of open and closed questions about the user experience, but also specific questions on game aspects in focus for the experiment, such as \textit{``Did you do this? Dig there? Press that button?''} Such specific questions can reveal how well the gamer has understood the game logic and mechanics.
Similar types of questions are used in interviews that allow for follow-up questions, which can provide richer answers and deeper insights into game play behaviour and experience. However, interviews are time consuming and can \textit{``take up too much time for ... sole developers''}. 
Balancing cost and benefit of experimentation poses a constant challenge. While experimenting with external participants takes time from development, it also reduces the risk of \textit{``running into problems that are very hard or costly to correct''}. 

\section {Pre-Release Development Stages}
\begin{figure}
    \centering
    \includegraphics[width=1\linewidth]{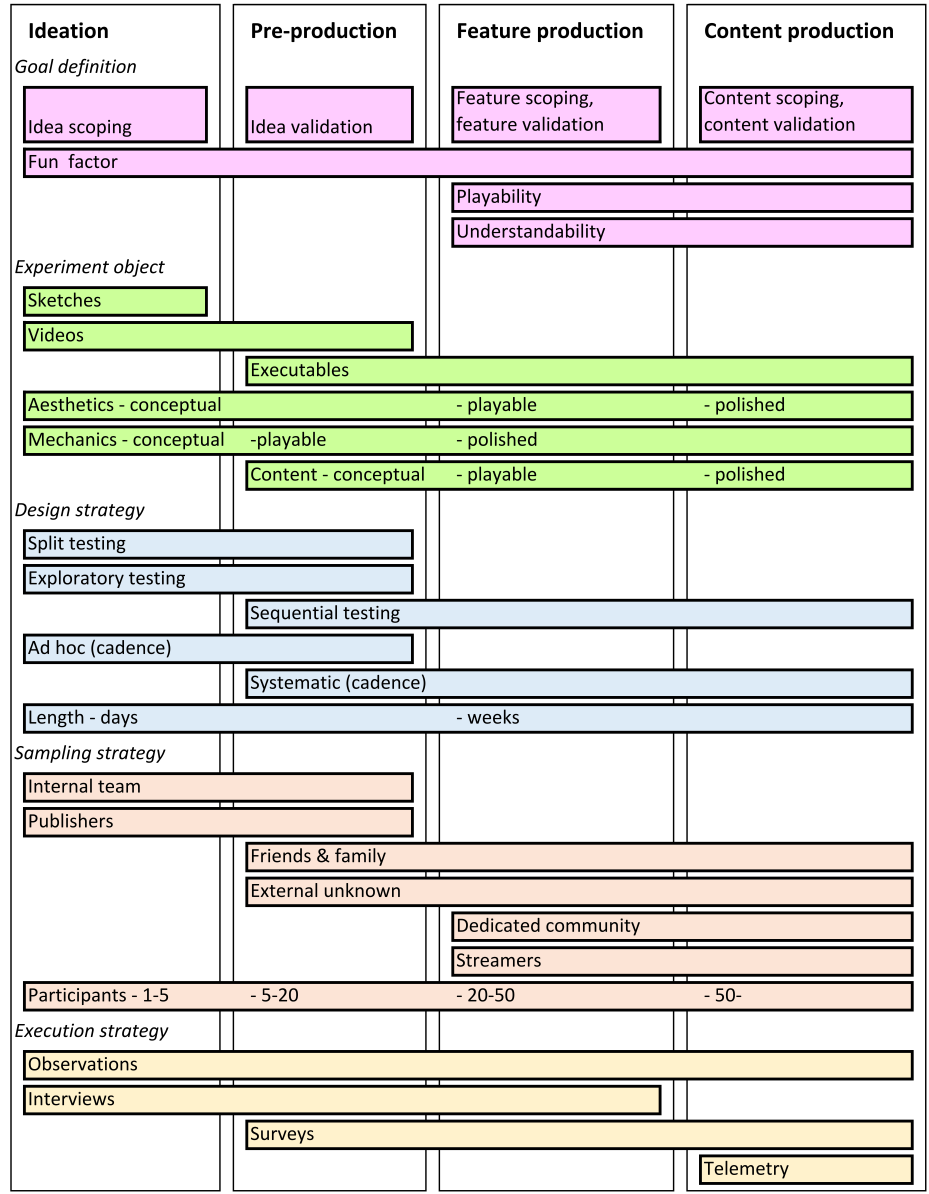}
    \caption{Overview of experimentation practices in pre-release development stages.}
    \label{fig:OverviewStages}
\end{figure}

Game development is an iterative and gradual process through which new ideas, features, and content of the game are explored, validated, refined, and expanded into a final public release. We have identified four main pre-release stages, namely \textit{Ideation}, \textit{Pre-production}, \textit{Feature production}, and \textit{Content production}. The experimentation practices for indie game development vary through-out these stages and experiments are carried out with a broadening set of participants and with a gradually extended and refined representation of the game from early sketches and designs to vertical slices through to alpha, beta, and final release. Fig. \ref{fig:OverviewStages} provides an overview of experimentation practices for all stages categorized using our CE framework (described in Section \ref{sect:CEFramework}.)

\subsection{Ideation Stage}
In this initial stage, the \textbf{goal} of experimentation is primarily to identify and \textit{scope the idea} of the game by validating its \textit{fun factor} and potential for market success. This stage is especially critical for Indie game companies since this affects the use of very limited development team resources. The game idea that is selected at this stage can determine the success or failure of the whole company. For example, showing a very early sketch to a publisher can lead to adjusting the scope and resulting in a game that \textit{``sells better in the end.''}

Simpler non-executable \textbf{experiment objects} are used to visualize a \textit{conceptual} representation of the envisioned \textit{aesthetics} and \textit{mechanics} of the game. This typically involves using \textit{sketches} and conceptual designs of the art and imagined gameplay to evaluate the game ideas and design options.
\textit{Videos} can also be used in this stage to demonstrate gameplay and to communicate an envisioned idea or game concept, e.g. to publishers and investors.

The \textbf{design strategy} in this stage often consists of \textit{ad hoc} experimentation using \textit{explorative testing} in short experiments that run for one or more days. \textit{Split testing} is also commonly used in this stage to explore multiple ideas in parallel, due to the relatively low cost of producing sketches and conceptual artwork. For example, several alternative versions of a game are often experimented with by showing conceptual art and designs
to potential publishers to present their game ideas and art designs.
The feedback is used to guide the indie game companies in selecting which idea to pursue and invest further development effort in. However, if several publishers are involved, this can lead to challenges with inconclusive feedback, which forces the indie game company to make decisions based on other criteria, e.g. lowest development cost.

\textbf{Sampling} of participants in this stage is limited to \textit{publishers} and the \textit{internal team}, and the \textbf{execution strategy} for experiments primarily relies on \textit{observations} and informal \textit{interviews}. At this stage, before the game idea and concept is set, it is very important for Indie game companies to include \textit{publishers} in the testing to ensure their support. 
For example, an experiment consisting of presenting conceptual sketches of different game ideas to a potential publisher can provide feedback that guides the decision on which idea to pursue and to develop further. This can prevent wasting scarce development resources on unviable game ideas.
Members from the \textit{internal team} are also important participants in experiments in this early stage. Sketches and videos can be used to visualise and convey the game idea and concepts, and support the development team in jointly discussing and shaping the game idea.


\subsection{Pre-Production Stage}
The main \textbf{goal} of the pre-production stage is to further develop and \textit{validate the idea} and to ensure that it is possible to develop it into a \textit{playable}, \textit{fun} and \textit{understandable} game for the target audience. In this stage, a lot of effort is put into developing the game mechanics and evaluating the \textit{playability} and \textit{fun factor} of the game. This can, and should, be done even before the game has nice-looking aesthetics, and ensure \textit{``that players enjoy it and actually have fun, even though the level is still ugly as hell''}. 

For games with a more complex game play, such as strategy games, the \textit{understandability} and learning aspects of a game should also be evaluated at this stage. In such games, players must understand the basic game mechanics to enjoy playing them, for example, which tools that can be used for a certain task and how. 

The \textbf{experiment objects} used for experimentation in this stage are primarily \textit{executable} prototypes with \textit{playable mechanics} and \textit{conceptual aesthetics} and \textit{content}. Even if these \textit{``early versions of the game ... aren't very refined, they work''} 
and allow game developers to get early feedback on their ideas. This feedback should be fed directly back into the game design and development, e.g. via the backlog.
In this stage, there is a strong focus on experimentation with the game mechanics. An example is an Indie first-person-shooter game where the main character is a non-human who neither moves or acts like a human, nor has hands to hold a gun with. In this case, early and frequent experimentation was performed to develop \textit{``the core mechanics... running around, shooting, reloading''}.
The main goal of this experimentation was to figure out how these new mechanics should work to provide the right feel and game experience. Placeholder-level art, e.g. created by an AI-based image generator, is a cost-effective means of experimentation that can be used to experiment and get the right \textit{feel} early on. 
Tutorials based on \textit{videos} can also be used at this stage, in particular to evaluate \textit{understandability} of more complex games, and how well players understand and learn to use the game mechanisms. 

Experimentation in this stage, primarily applies a \textbf{design strategy} consisting of \textit{systematic experimentation} through a combination of methods over small and quick iterations, typically 1-2 weeks.
\textit{Split} and \textit{exploratory} testing are used as the game mechanics are evolved, though for different purposes. \textit{Split-testing} is useful for experimenting with new features. As one Indie game developer described their use of A/B testing in this stage, \textit{``we build a quick and dirty version of the one and the other and let them test.''}
One indie company experimented with advanced AI functionality by developing two versions of their game, one with the AI functionality. In the experiment, the participants selected which version they preferred to play. 
Based on the feedback, the company decided to adopt the new AI functionality.
Testing multiple versions at the same time also comes with challenges, especially for indie companies who have access to a limited number of players. Questions arise, such as \textit{``should they all play both versions? And if so, in what order? And does it influence them?''}
The sampling of participants and planning of the experiment then needs to be considered to mitigate biases and reduce the risk of incorrect or inconclusive feedback.


A \textbf{sampling strategy} based on convenience and availability tends to be used in the pre-production stage, and experimentation is primarily carried out with participants from the \textit{internal team} and the \textit{publishers}. A small set of easily available participants from within the teams co-working space can also be used as playtesters. Internal experimentation is often performed on a weekly basis, e.g. through \textit{show-and-tell sessions} that then continue also throughout later development stages. At the end of this stage, as the prototype becomes more mature, early versions of the game can be shown to \textit{friends \& family} and \textit{external participants} can be recruited through publishing small demos as previews on gaming platforms such as Itch.io and Steam.
These kinds of experiments can be used to poll interest and can provide valuable \textit{``feedback from people who explore the game''}.
However, experimenting with such early versions of the game with external gamers can also yield \textit{``very divided feedback from enthusiastic to very disappointed.''} Drawing conclusions from such diverse feedback can be very challenging and should not be used as the sole source of feedback for major decisions about the game.

\subsection{Feature Production Stage}
The main \textbf{goal} of experimentation turns to \textit{feature scoping} and \textit{feature validation} as development shifts into developing a complete set of features and working towards an alpha release. In this stage, \textit{``the decision on what to include in the game''} 
is based on experimentation results, both regarding which features to include and the detailed scoping of these. As the game matures, all aspects of the game are important to consider and validate including \textit{fun factor}, \textit{playability} and \textit{understandability}.

The \textbf{object} that is experimented with is primarily a version of the game under development. Experimentation is then perform with an \textit{executable} that has \textit{playable aesthetics} and \textit{content}, and gradually \textit{maturing mechanics}.

The predominant \textbf{design strategy} in this stage is \textit{sequential} testing using play testing in a \textit{systematic} manner over a period of 1-2 \textit{weeks} with predefined hypothesis or questions for the playtest. Lessons and takeaways from experimentation are used to inform the backlog and prioritise feature development and bug fixing. An example
is experimentation with replacing a gun with an alternative mechanisms. In this case, the goal of the experiment was to evaluate a hypothesis that the gun was used too repetitively and impacted negatively on the user experience. In this case, the experiment revealed that the users enjoyed, and preferred, the gun, which led to re-implementing it in the game. \textit{Split testing} is also used in this stage when experimenting with specific new features, but to a limited degree due to the cost of implementing multiple options. An example of a split test is an experiment where alternative \textit{playable game mechanics} and \textit{playable aesthetics} in the form of alternative artwork were added to a limited prototype based on the main game. In this case, the alternative variant was compared to the main game and the ideas that were successful according to the participants' feedback, were re-implemented in the main game.

The \textit{length} of experiments typically shifts to longer iterations in this stage, as more development is conducted between tests, and the maturity level increases for the mechanics and aesthetics of the game. This is sometimes mixed with shorter and more intense periods of experimentation. Shifting the \textit{cadence} allows for both experimenting with more strategic features which require \textit{``more long-term decisions and introduction of larger feature implementations, followed by more intense testing''} of smaller changes to optimize the game mechanics or details in a feature.

In this stage, the \textbf{sampling} of participants shifts to involve a larger number of \textit{external unknowns}, while the \textbf{execution strategy} remains similar to pre-production with an increase in \textit{surveys} and recorded \textit{observations} of gameplay. 
Indie game companies often recruit external participants by publishing early game versions, or demos, on  platforms such as Itch.io and Steam to attract attention and to invite players into their Discord channels (a chat platform). The Discord channels serve as an infrastructure for growing and facilitating \textit{dedicated communities} of fans and potential playtesters that can be used in experiments on a reoccurring basis. If Indie companies have the funds, they can also acquire external playtesters from third party platforms. These platforms provide recruitment of game players with specific characteristics. While experimenting with 'the correct' audience reduces the risk of biases in the outcome, it also comes with a monetary cost. Another, cheaper, source of valuable insights into game play can be obtained by observing \textit{streamers} and their audience,
or by letting \textit{friends \& family} evaluate early demos of the game.

\subsection{Content Production Stage}
As development shifts to working towards a beta release, the main \textbf{goal} of experimentation turns to \textit{content scoping} and \textit{content validation} in the content production stage. Developing content can be seen as \textit{``stretching out that fun and putting these features in varied situations''} 
and \textit{``making actual story levels that use all of these features and tools that we've built, but... [with] variations .. so that the game stays interesting.''}
This requires a lot of experimentation to find out \textit{``which combinations of mechanics that work and are interesting.''}

Similarly to the feature production stage, the main \textbf{experiment object} in this stage is the game under development, which at this point is an \textit{executable} with a full set of features with \textit{polished mechanics}. The \textit{aesthetics} of the game are primarily conceived and evaluated in this stage through producing the final artwork, and developing this game aspect from a \textit{conceptual} to a \textit{polished} level of maturity.

A \textit{sequential} approach is the predominant \textbf{design strategy} currently used by indie companies in the content production stage. Level designers design and create game situations that are then experimented with, and  \textit{``if people actually struggle there, we'll try the other [alternative] next.''}

The \textbf{sampling} of participants  shifts as the game matures and play testing is performed with larger numbers of \textit{external unknowns} and \textit{dedicated community}. These are complemented by continuous game testing with the \textit{internal team} since 
play testers with game development experience can smoke out issues that normal gamers may not observe. The opposite is also true, an external game testers may finds issues and bugs that the internal team are blind to.

The number of participants allows shifting the \textbf{execution strategy} towards increased use of \textit{telemetry}, meaning data generated by players' interactions with the game. Telemetry data, such as duration and progress of game play, can provide insights about game play including difficulties, engagement, bugs, etc.. In the mobile context, game events are tightly connected to analytics, e.g., to record time from one game action to another to tell how quickly a player progresses compared to an average user.
Frameworks such as Google Analytics can provide certain metrics such as the number of active users, new players, monthly active users, day-one retention, and day-seven retention. While these general KPIs are helpful, more detailed in-game metrics can provide more value. However, this requires indie companies to implement analytics tooling such as data pipelines and processing infrastructure of the back-end, in their development environment.
However, the \textit{telemetry data} needs to be complemented by \textit{interviews} and \textit{observations} in order to evaluate the fun factor since \textit{``fun is so subjective''}.
For example, observing how players move around and interact with the game environment provides insights into the game experience, how the game mechanics are working, and how much \textit{fun} the participants are having. 
This qualitative feedback is especially time consuming to gather and to analyse and game developers should strive for a good cost-benefit balance, e.g. by limiting this data collection to a subset of the participants of an on-line experiment.

\section{Recommendations and Advice for Practice}
We provide the following recommendations and advice to practitioners on how to improve on their experimentation practices. These are based on our empirical insights and on previous research on continuous experimentation.

\textbf{Adopt an experiment-driven mindset} based on continuous experimentation with game ideas, features, and content. Be prepared to pivot and change direction, and even kill favourite game ideas, if experimentation results indicate that they are unviable.

\textbf{Systematically and consciously perform experimentation} throughout the game development effort, from ideation, through pre-production, to production and beyond. Define hypothesis and clear goals for each experiment, and design the experiments and the evaluation factors to match this.

\textbf{Mix and combine different types of experiments} to maximise the value of the feedback and reduce the risk of inconclusive or incorrect conclusions. By varying and combining the strategy for sampling participants, and designing and executing experiments, multiple triangulating data points can be obtained, which increase the reliability and validity of the data.

\textbf{Involve external participants} as early as possible, ideally already in the ideation stage. Feedback from external unknowns and dedicated community participants often provides un-biased feedback on new game ideas and mechanics, in contrast to the feedback from internal testers and friends \& family. Such feedback enables detecting problems early on and enables efficient development of features and content with a high game experience factor. This avoids spending development resources on major changes in later development stages.

\textbf{Select the execution strategy to match available number of participants}. Focus on observations and interviews when you have a limited number of participants. And, use telemetry data when you have access to a larger numbers of participant. We also recommend complementing telemetry data with observations and interviews for a smaller number of participants to provide insights into how to interpret the quantitive data.

\textbf{Adapt your design strategy} to the current design stage of the item under development, and the overall game development stage and the number of available participants. In early design stages, e.g. of a new feature, we recommend experimenting with multiple variants through split testing, if there are sufficient participants for this and prototypes can be produced cheaply. For example, experimenting with multiple variants of sketches (for content and aesthetics) and smaller executable demos can provide a cost effective initial evaluation of multiple design options in a short period of time. As the design of a feature matures, or if the cost of producing multiple variants is very high, sequential testing is the most cost effective choice. Explorative testing is recommended as a complement to the other testing strategies, in particular for later development stages and of later design stages, when the feature under development is mature. 

\textbf{Be aware of and mitigate biases} by using mixed experimentation strategies. For example, by varying the data collection methods, the number and type of participants used, the cadence and execution strategy of experiments. For example, combine telemetry with observations and interviews to gain explanations to the measured behaviour. The risk of drawing incorrect conclusions can be reduced both by combining different types of data and data from multiple sources. Such triangulation can also mitigate the risk of biases and increase the value obtained from experiments.

Finally, we recommend \textbf{continuously improving experimentation practices} through regular retrospectives of performed experiments to capture insights gained from specific experiment designs and set-ups. Our framework can serve as a basis for team discussions on how to improve on experimentation practices. Questions to consider include: Could similar value have been obtained with a simpler and cheaper prototype (Experiment object)? How did the number and type of participants (Sampling strategy) affect the cost and outcomes? Was the data collection method appropriate (Execution strategy)? Also, consider the current development stage and reflect on how experimentation can be tailored to provide the most valuable feedback and insights for that stage.

\section{Conclusions}

Indie companies face significant challenges due to limited time and resources compared to larger and more established game companies with well-known game titles. To succeed, these startups must adopt a cost-benefit mindset and effective experimentation practices. Our research highlights the importance of continuous experimentation in indie game development and we identify key aspects such as goal definition, experiment object, design strategy, sampling strategy, and execution strategy. We also describe current experimentation practices for the pre-release development stages of ideation, pre-production, feature production and content production. Experimentation is crucial in pre-release stages to scope and validate the business viability of game ideas, features, and content. By continuously evaluating the game experience through experiments with external gamers (users), problems with playability, fun factor, and understandability can be caught early on. The feedback can be used to guide design decisions and enable effective use of scarce development resources.

Future work will delve deeper into the key aspects of experimentation practices and factors influencing them in order to identify best practices in this area and means to further improve game development practices. Special attention will be given to the resource and efficiency needs of indie game companies and how to support game developers in optimising their cost-benefit balance. This involves investigating tools and processes for experimentation, and exploring and validating the usefulness of our recommendations. Additionally, the perspectives and practices of more established and larger game companies will be explored to contrast with the findings from indie game companies. 

Finally, our findings and recommendations can provide insights and practical advice on how to apply experimentation practices. We encourage game developer both within indie companies and more established game companies to consider our findings and how to strengthen their experimentation practices in the pre-release stages of game development. Also other software-intensive industries, in particular, those with a strong focus on user experience may gain from our considering our advice. 

\section*{Acknowledgments}
[Anonymised]
\bibliographystyle{splncs04}
\bibliography{mybibliography}

\end{document}